\documentclass[aps,prd,superscriptaddress,showpacs,showkeys,preprintnumbers,nofootinbib]{revtex4}
\usepackage{graphicx,color}
\usepackage{amsmath}
\usepackage{amsfonts}
\usepackage{amssymb}

\newcommand{\Tr}{\rm Tr} 
\newcommand{\half}{ {\textstyle\frac{1}{2}} }
\newcommand{\eps}{\epsilon}
\newcommand{\cdott}{{\mskip -1.5mu} \cdot {\mskip -1.5mu}}
\newcommand{\ii}{i}                    
\newcommand{\de}{d}                    

\begin{document}

\title{Electron Ion Collider transverse spin physics\footnote{Talk presented at International Workshop on Diffraction in High-Energy Physics, Otranto, Italy, September 10 - 15, 2010}}

\author{Alexei Prokudin}
\email{prokudin@jlab.org}
\affiliation{Jefferson Lab,12000 Jefferson Avenue, Newport News, VA 23606}

\begin{abstract} Electron Ion Collider is a future high energy facility for 
studies of the structure of the nucleon. Three-dimensional parton structure is one of the main goals of EIC. In momentum space Transverse Momentum Dependent Distributions (TMDs) are the key ingredients to map such a structure.
At leading twist spin structure of spin-1/2 hadron can be described by 8 TMDs.  Experimentally these functions can be studied in polarised SIDIS experiments.
We discuss Sivers distribution function that describes distribution of unpolarised quarks in a transversely polarised nucleon and transversity that measures distribution of transversely polarised quarks in a transversely polarised nucleon. 
\end{abstract}

\pacs{13.88.+e, 13.60.-r, 13.15.+g, 13.85.Ni}
\keywords{Transverse Momentum Dependent distributions, Spin Asymmetries, Semi Inclusive Deep Inelastic Scattering}

\maketitle


Electron Ion Collider (EIC) is proposed by EIC Collaboration \cite{EIC} that includes more than 100 physicists from over 20 laboratories and universities.

EIC at medium -- high energy of $\sqrt{s} \sim 20 \div 70$ GeV will allow high precision measurements with polarised proton and  ion H, D, He$^3$, Li beams. 
Luminosity will be $\simeq 10^{34}$ sm$^{-2}$s$^{-1}$ which is comparable to that of fixed target experiments and is unprecedented for a collider.

Three-dimensional parton picture of the nucleon is one of the main goals of an EIC and such a three-dimensional distribution can be mapped by studying Transverse Momentum Dependent distribution functions (TMDs). TMDs depend on two independent variables: fraction of hadron momentum carried by parton, $x$, and intrinsic transverse momentum of the parton, ${\bf p}_T$. 

The quark-quark distribution correlation function is defined
as \cite{Mulders:1995dh,Bacchetta:2006tn}
\begin{equation}  
{\Phi_{ij}} 
(x,{\bf p}_T)= \int 
        \frac{\de \xi^- \de^2 {\bf \xi}_T}{(2\pi)^{3}}\; 
 e^{\ii p \cdot \xi}\,
       \langle P, S_P|\bar{\psi}_j(0)\,
{\cal U}^{n_-}_{(0,+\infty)}\,
{\cal U}^{n_-}_{(+\infty,\xi)}\,
\psi_i(\xi)|P ,S_P\rangle \bigg|_{\xi^+=0}
\label{e:phi} 
 \end{equation}   
with $p^+ = x P^+$, $P^+ = {(P^0+P^3)}/{\sqrt{2}}$ is the big component of proton's momentum. ${\cal U}$ is the gauge link (Wilson line) that assures color gauge invariance of the correlator ${\Phi_{ij}}$.

Corresponding distribution functions can be obtained by projecting the correlator onto the full basis of $\gamma$ matrices $\Phi^{[\Gamma]} = \half \Tr [
\Phi\, \Gamma]$. At leading twist (expansion in $P^+$) the spin structure of the proton can be described by 8 TMDs (see \cite{Mulders:1995dh,Bacchetta:2006tn} and references therein):
\begin{eqnarray} 
\label{e:1}
\Phi^{[\gamma^+]}(x,{p}_{T}^{2}) & = &
 f_1
(x,{p}_{T}^{2}) 
 - \frac{\eps_{T}^{\rho\sigma} p_{T \rho}^{}S_{T \sigma}^{}}{M} \, 
   f_{1T}^{\perp}
(x,{p}_{T}^{2}) 
 \,,
\\
\Phi^{[\gamma^+ \gamma_5]}(x,{p}_{T}^{2}) & = &
 S_L \, g_{1L}
(x,{p}_{T}^{2}) 
 - \frac{{p}_{T} \cdott {S}_{T}}{M} \, 
   g_{1T}
(x,{p}_{T}^{2}) 
 \,,
\\ 
\Phi^{[\ii \sigma^{\alpha +}\gamma_5]}(x,{p}_{T}^{2}) & = &
 S_{T}^{\alpha} \, h_{1}
(x,{p}_{T}^{2}) 
 + S_L\,\frac{p_{T}^{\alpha}}{M} \, h_{1L}^{\perp}
(x,{p}_{T}^{2}) 
\nonumber \\   & & \hskip -0.4cm
 - \frac{p_{T}^{\alpha} p_{T}^{\rho}
     -\frac{1}{2}\,{p}_T^{2}\,g_T^{\alpha\rho}}{M^2}\, S_{T \rho} 
   h_{1T}^{\perp}
(x,{p}_{T}^{2}) 
- \frac{\eps_{T}^{\alpha\rho} p_{T \rho}^{}}{M}  
   h_{1}^{\perp} 
(x,{p}_{T}^{2}) 
\end{eqnarray}
here the projections are done with respect to the quark's polarisation,
$\Phi^{[\gamma^+]}$ is distribution of unpolarised quarks, $\Phi^{[\gamma^+ \gamma_5]}$ is distribution of longitudinally polarised quarks and $\Phi^{[\ii \sigma^{\alpha +}\gamma_5]}(x,{p}_{T}^{2})$ is the distribution of transversely polarised quarks.

Experimental data are available from fixed target SIDIS $\ell P(S_P) \rightarrow \ell' h X$ experiments HERMES and COMPASS.
BELLE experiment provides data on $e^+e^- \rightarrow h_1 h_2 X$ annihilation.
RHIC has data on spin asymmetries in hadron production $P(S_P) P \rightarrow h X$. Future experimental data ah high-$x$ will come from JLab. 

\begin{figure}{}
\hskip 0.5cm\includegraphics[width=.25\textwidth,bb= 10 140 540 660,angle=-90]{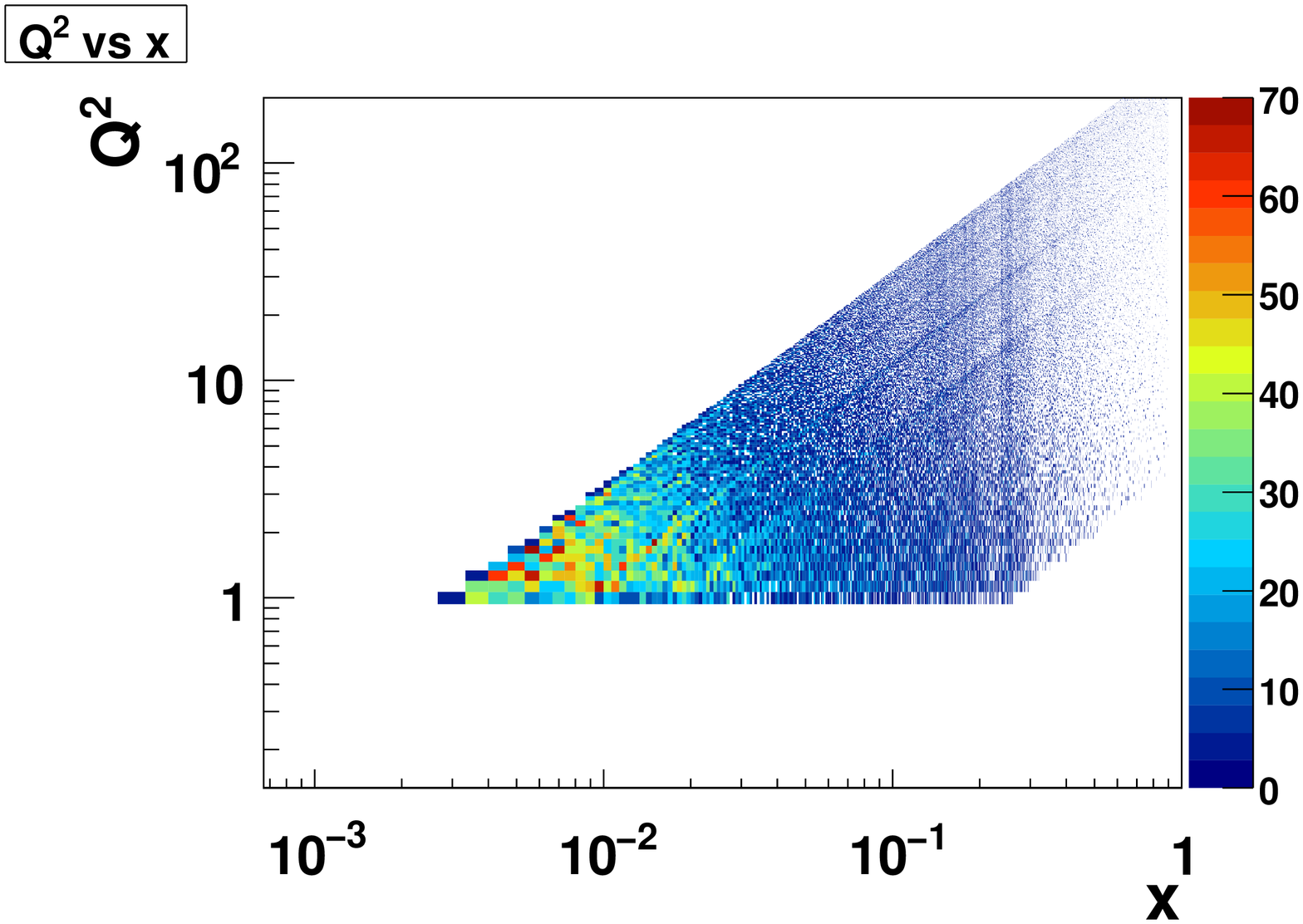}
\hskip 2.5cm
\includegraphics[width=.25\textwidth,bb= 10 140 540 660,angle=-90]{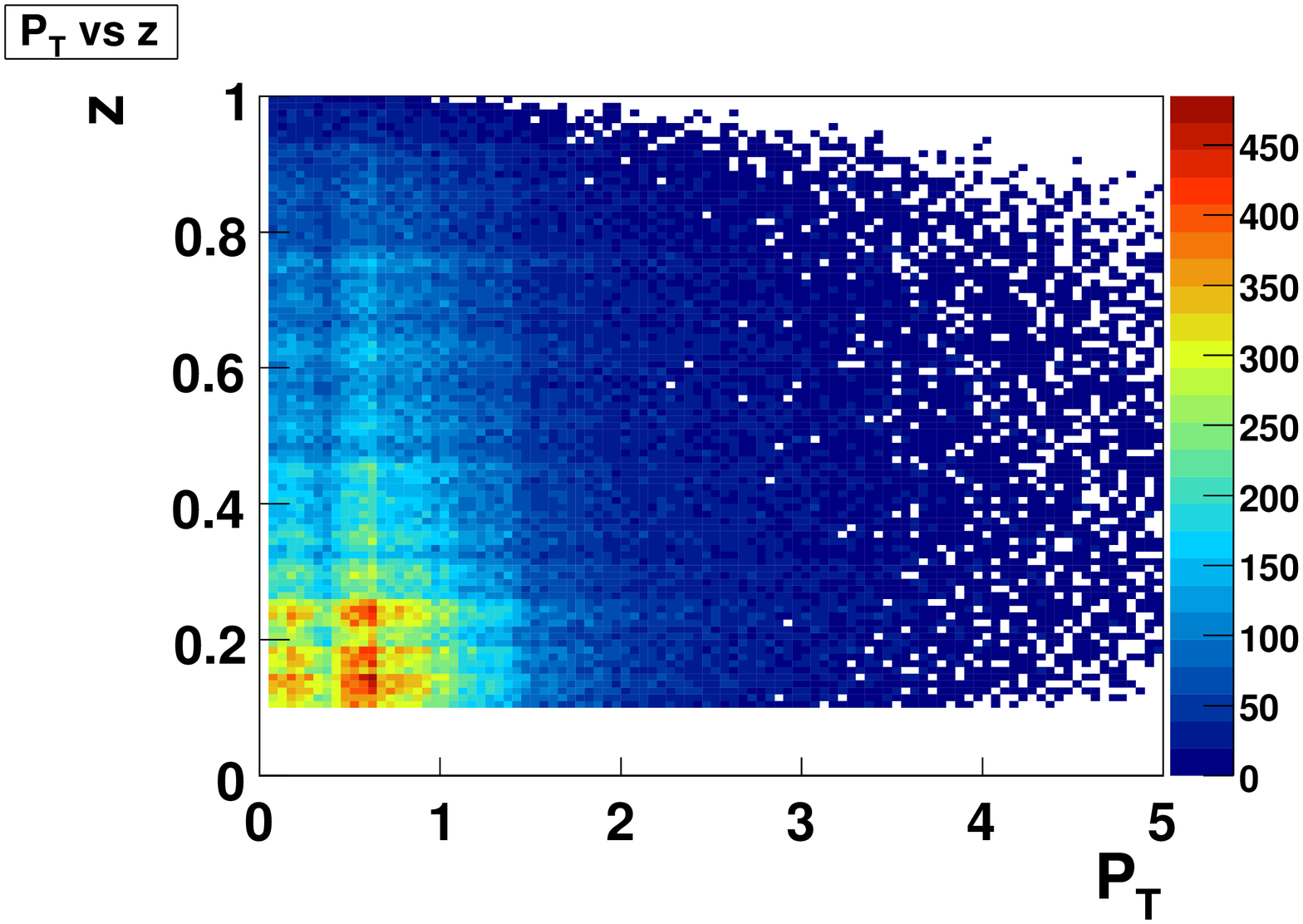} \\
\caption{Kinematical coverage in $x-Q^2$ and $z-P_{h\perp}$ of EIC at $\sqrt{s}=20$ GeV.}
\label{fig_EIC}
\end{figure}

One of the main advantages of EIC is a wide kinematical region in $x-Q^2$ and $z-P_{h\perp}$ variables, see Fig.~\ref{fig_EIC}. The $Q^2$ range will allow us a unique opportunity to study evolution of TMDs. Low-high $P_{h\perp}$ range is a possibility to study inteplay of TMD~\cite{Ji:2004wu} and collinear~\cite{Collins:1989gx,Qiu:1991pp} factorization schemes.

Electron Ion Collider will allow us to study sea quark and gluon TMD distributions and $Q^2$ range will allow us to study evolution of TMDs.

\noindent {\bf Transversity} \\
Transversity distribution $h_1(x,{p}_{T}^{2})$ \cite{Ralston:1979ys} describes distribution of transversely polarised quarks inside a transversely polarised nucleon. Tensor  charge measures net transverse polarisation of quarks
\begin{equation}
\delta_T q = \int_0^1 dx( h_{1q}(x) - h_{1\bar q}(x) ).
\end{equation}

In SIDIS transversity can be measured together with chiral-odd fragmentation function $H_{1q}^\perp (z,{k}_{T}^{2})$ so-called Collins fragmentation function \cite{Collins:1992kk}
\begin{equation}
A_{UT}^{\sin (\phi_h +\phi_S)} \propto \frac{\sum_q e_q^2\, h_{1q} \otimes H_{1q}^\perp}{\sum_q e_q^2\, f_{1q} \otimes D_{1q}}\; ,
\end{equation}
where $\phi_h$ and $\phi_S$ are azimuthal angles of produced hadron and target polarisation vector with respect to lepton scattering plane.

Experimental data on SIDIS asymmetries $A_{UT}^{\sin(\phi_h + \phi_S)}$ from HERMES, COMPASS collaborations \cite{Diefenthaler:2007rj,Alekseev:2008dn} and $e^+ e^- \to h_1
h_2 X$ data from BELLE collaboration \cite{Seidl:2008xc} allow extraction\cite{Anselmino:2007fs,Anselmino:2008jk} of transversity and Collins fragmentation function.

\begin{figure}{}
\hskip 0.5cm\includegraphics[width=.25\textwidth,bb= 10 140 540 660,angle=-90]{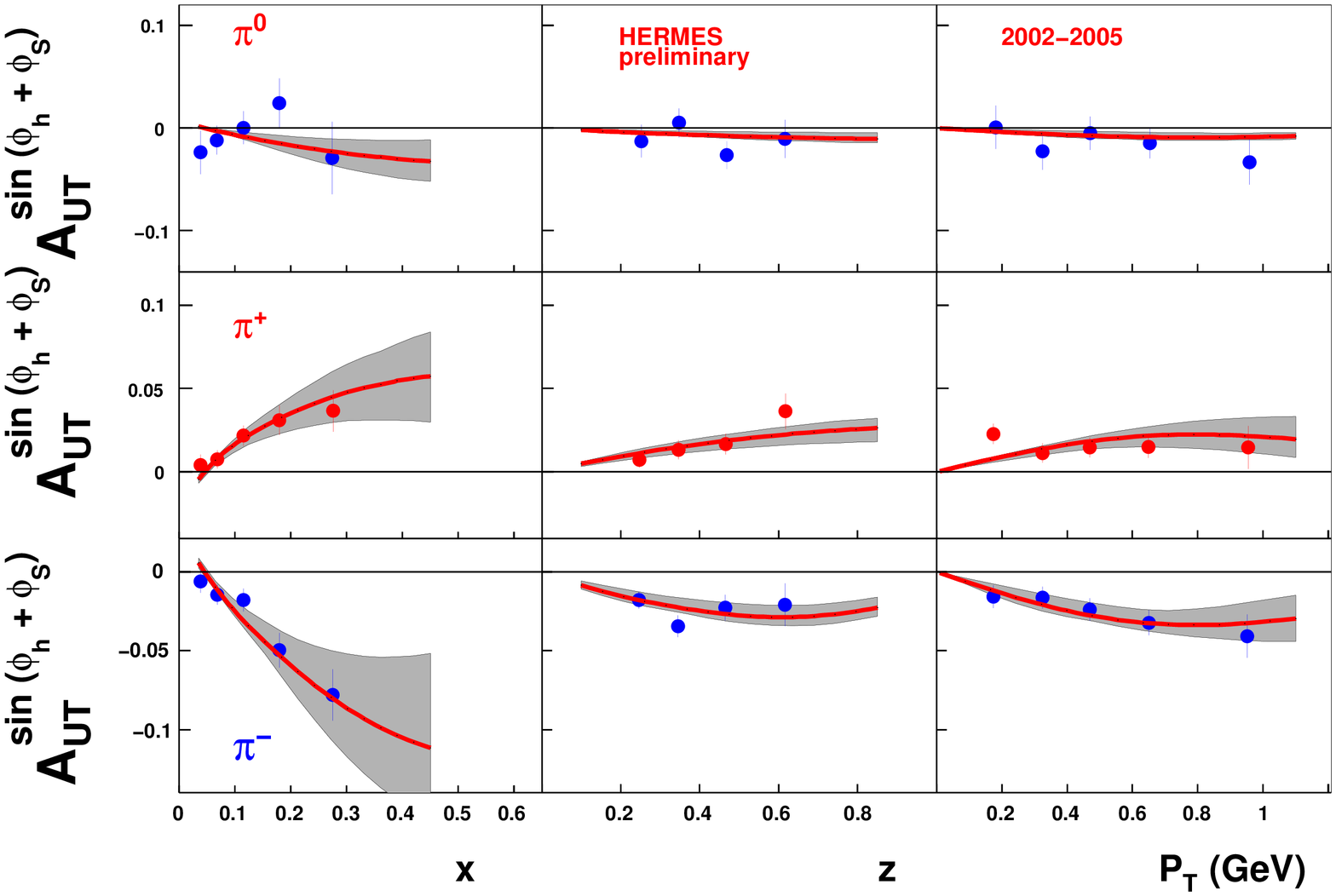}\hskip 2.5cm
\includegraphics[width=.25\textwidth,bb= 10 140 540 660,angle=-90]{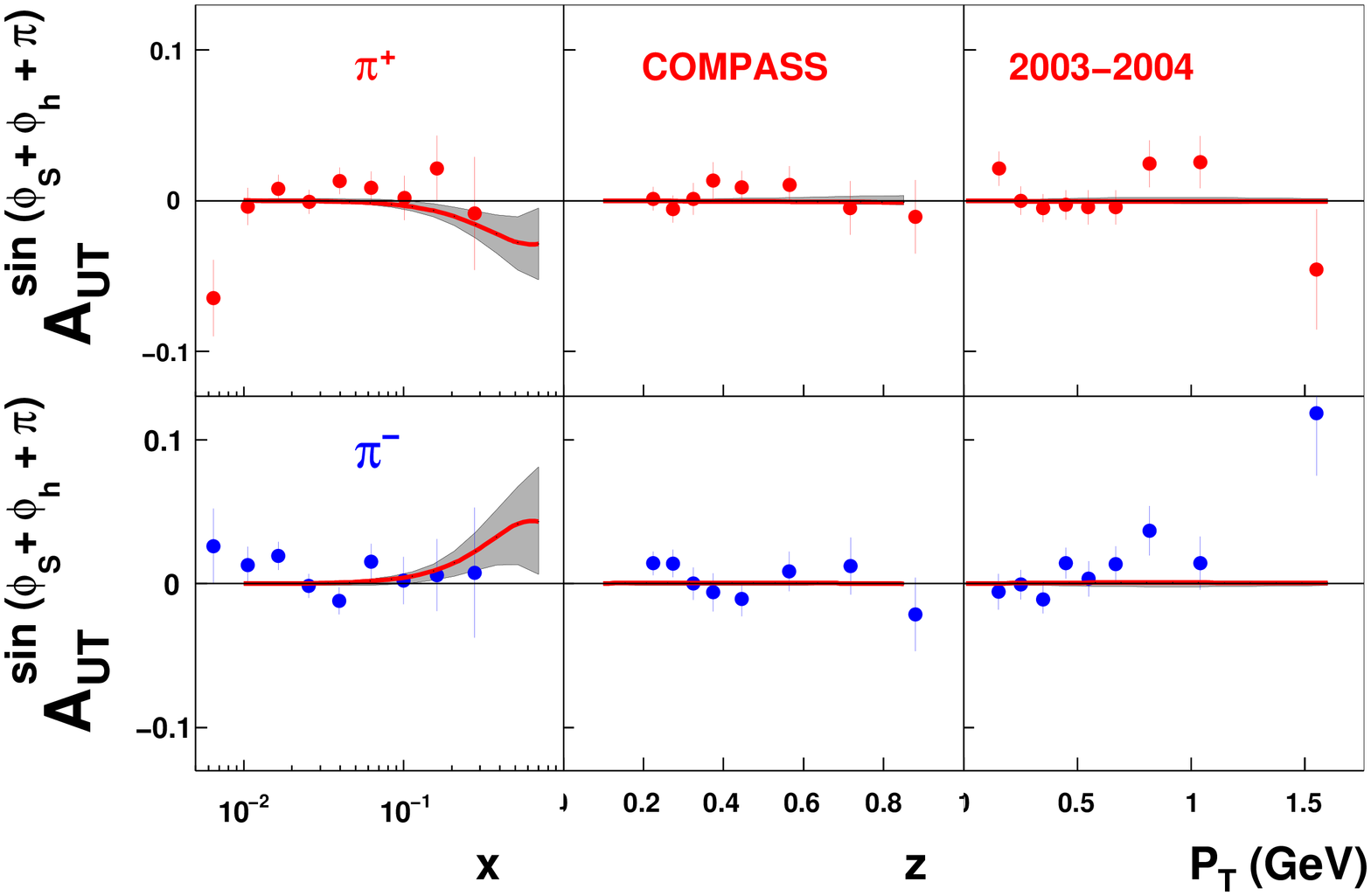} \\
\caption{Fit\cite{Anselmino:2008jk} of HERMES \cite{Diefenthaler:2007rj} (left panel) and COMPASS \cite{Alekseev:2008dn} data (right panel).}
\label{fig_0}
\end{figure}

\begin{figure}{}
\hskip -3.5cm \includegraphics[width=.2\textwidth,bb= 29 15 570 660]{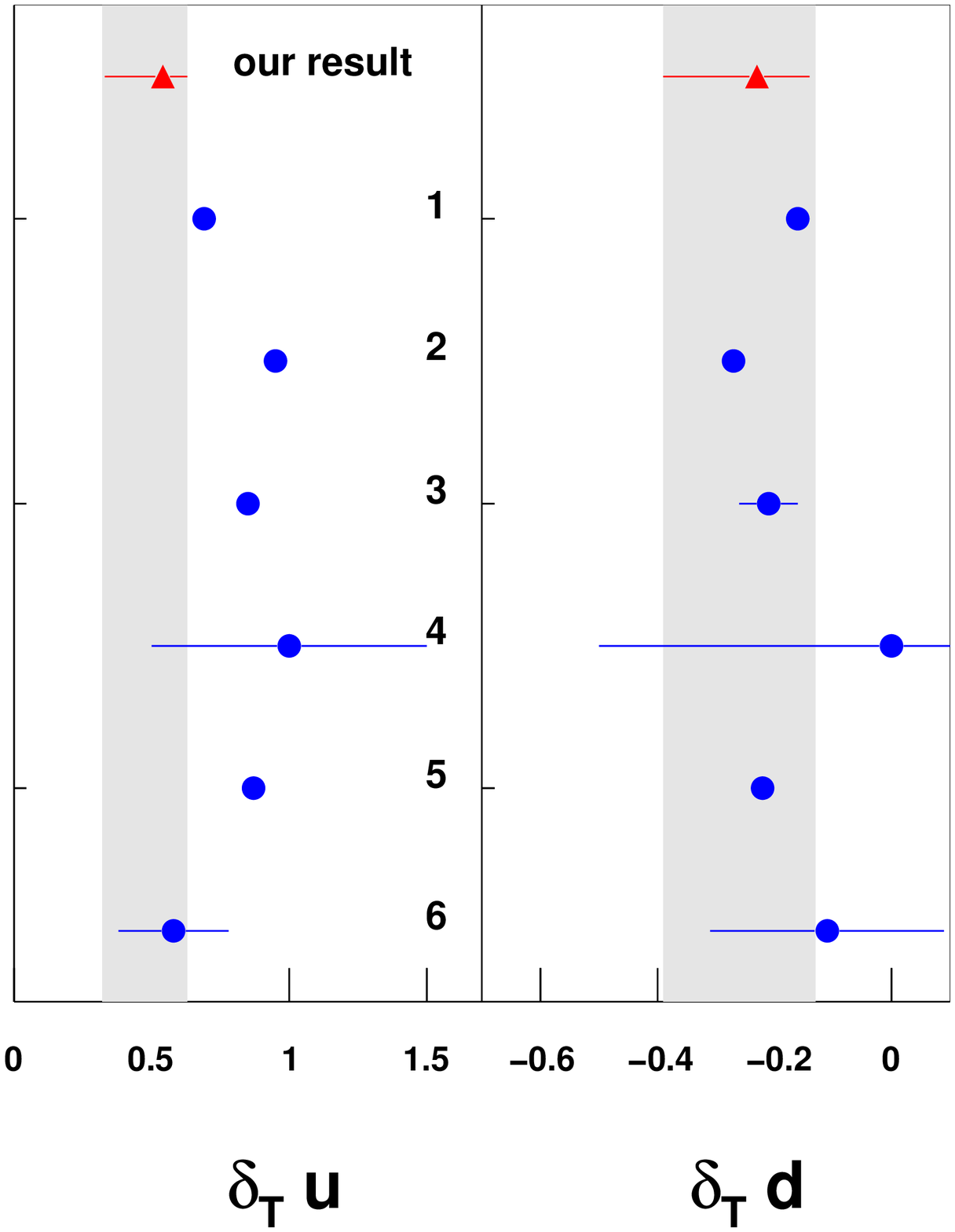} \hskip 2.cm 
\includegraphics[width=.25\textwidth,bb= 535 50 76 670,angle=90]{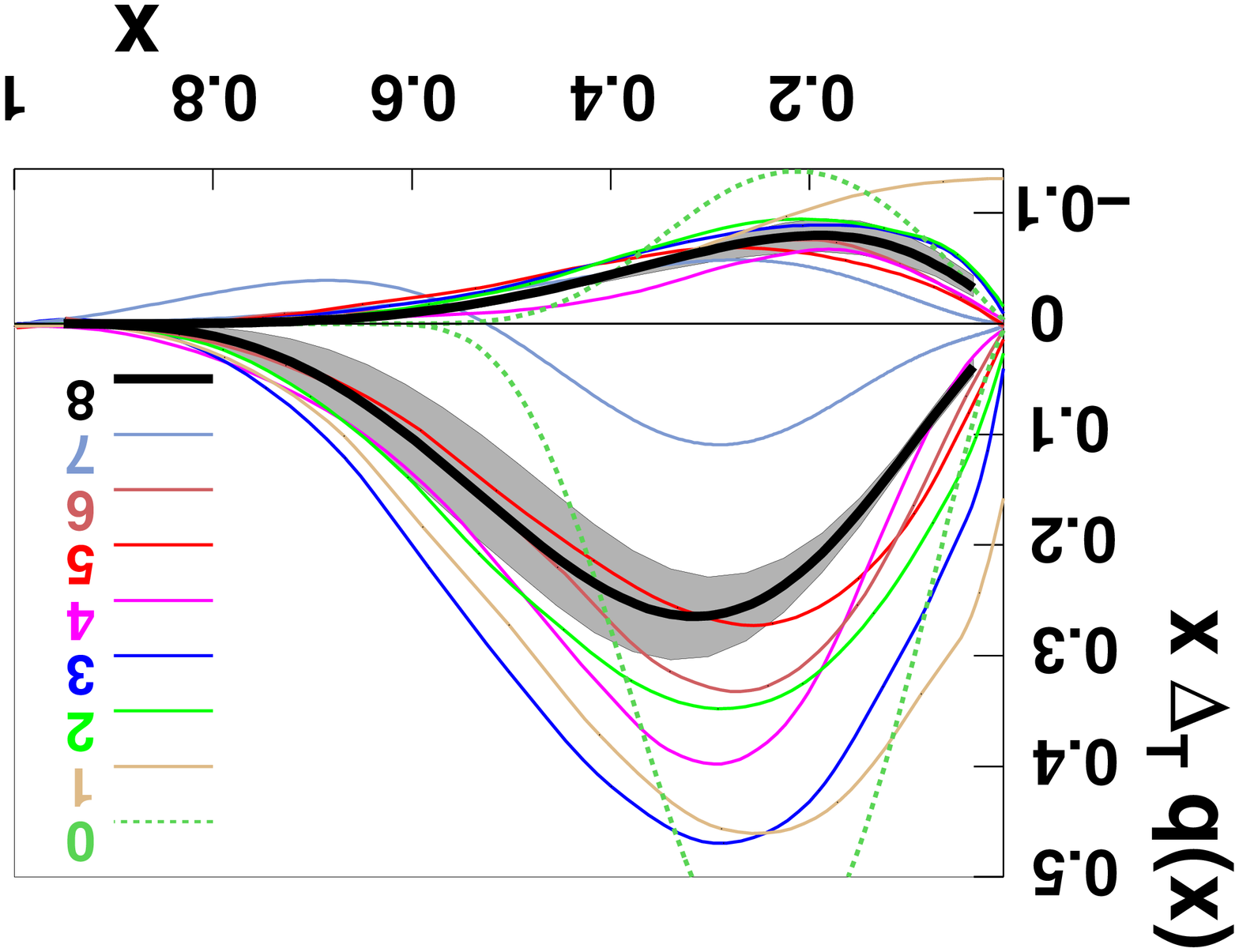}
\caption{Left panel: extraction of tensor charge \cite{Anselmino:2008jk} compared to models \cite{Cloet:2007em,Wakamatsu:2007nc,Pasquini:2005dk,Gockeler:2005cj,He:1994gz,Gamberg:2001qc}. Right panel: transversity for $u$ and $d$ quarks \cite{Anselmino:2008jk} compared to models \cite{Barone:1996un,Ma:2001rm,Schweitzer:2001sr,Pasquini:2005dk,Bacchetta:2008af,Cloet:2007em,Wakamatsu:2007nc}}
\label{fig_1}
\end{figure}

In Fig.~\ref{fig_0} we show the best fit to the HERMES \cite{Diefenthaler:2007rj} and COMPASS \cite{Alekseev:2008dn} data, respectively.

We include one of the two sets of Belle \cite{Seidl:2008xc} data either $A_{12}$ or $
A_0$ data, see for details Refs.\cite{Anselmino:2007fs,Anselmino:2008jk}. In this analysis we use $A_{12}$ data, the $\cos(\phi_1 + \phi_2)$
method.

\begin{equation}
A_{12}(z_1,z_2,\theta,\varphi_1 + \varphi_2) \propto
\cos(\phi_1 + \phi_2) \frac{\sum_q
e^2_q \, H_{1 q}^\perp (z_1)\,
 H_{1 \bar q}^\perp (z_2)}{\sum_q e^2_q \, D_{1 q}(z_1)
 D_{1 \bar q}(z_2)}\, ,
\end{equation}
here $\phi_1$ and $\phi_2$ are azimuthal angles of produced hadrons $h_1$ and $h_2$ with respect to the plane formed by the quark thrust axis and electron-positron pair, in a so-called Collins-Soper frame. 

The extracted values \cite{Anselmino:2008jk} of tensor charge are
$\delta_T  u = 0.54^{+0.09}_{-0.22}$,  $\delta_T  d = -(0.23^{+0.09}_{-0.16})$
at $Q^2 = 0.8$ GeV$^2$.

In Fig.~\ref{fig_1} we plot tensor charge and transversity distributions compared to models. As can be seen from Fig.~\ref{fig_1} the experimental precision is still not good enough to discriminate among models. Anti-quark transversity distributions and $Q^2$ dependence of asymmetry is not yet known experimentally. JLab data will allow to expand $x$ range of transversity extraction.  
 Electron Ion Collider will be able to shed light on anti-quark distributions.

\noindent {\bf Sivers distribution function} \\
$f_{1T}^{\perp}
(x,{p}_{T}^{2})$ is so-called Sivers function \cite{Sivers:1989cc}, it describes correlation between orbital angular motion of quarks and the spin of the proton $\eps_{T}^{\rho\sigma} p_{T \rho}^{}S_{T \sigma}^{}$. This function exists due to the presence of Final State Interactions of the struck quark and the remnant of the nucleon after the interaction.

Distribution of unpolarised quarks inside a transversely polarised nucleon can be written as
\begin{equation}
f_{q/p^\uparrow} (x,{\bf p_T}) = f_{1} (x,p_T^2) -
f_{1T}^{\perp}(x,p_\perp^2)
\frac{{\bf S} \cdot (\hat {\bf P}  \times
{\bf p_T})}{M} \; ,
\end{equation}
here ${\bf S} \cdot (\hat {\bf P}  \times
{\bf p_T})$ is correlation between quark motion and the spin of the nucleon. The function that measures this correlation is Sivers function $f_{1T}^{\perp}$.

Sivers function is a natural candidate for ``golden'' observable at EIC as it incorporates the main features of TMD physics:
it allows three-dimensional mapping of partons inside of the nucleon, it is related to spin-orbit correlations and gives access to Orbital Angular Momentum of partons and it represents non trivial color dynamics of QCD.

This function obeys a modified universality, it changes sign from SIDIS to Drell-Yan \cite{Collins:2002kn} process
$
f_{1T}^{\perp}
(x,{p}_{T}^{2})^{SIDIS} = - f_{1T}^{\perp}
(x,{p}_{T}^{2})^{DY},
$
the prediction of change of sign based on color gauge symmetry and parity and time reversal invariance $\mathcal{P},\mathcal{T}$ of strong interactions. Experimental test of this relation is very important for our understanding of QCD.

The asymmetry associated with Sivers effect is
\begin{equation}
A_{UT}^{\sin (\phi_h -\phi_S)} \propto \frac{\sum_q e_q^2\,  f_{1T}^{\perp}\otimes D_{1q}}{\sum_q e_q^2\, f_{1q} \otimes D_{1q}}\;.
\end{equation}

\begin{figure}{}
\centering
\hskip -0.5cm\includegraphics[width=.35\textwidth,bb= 25 50 740 550]{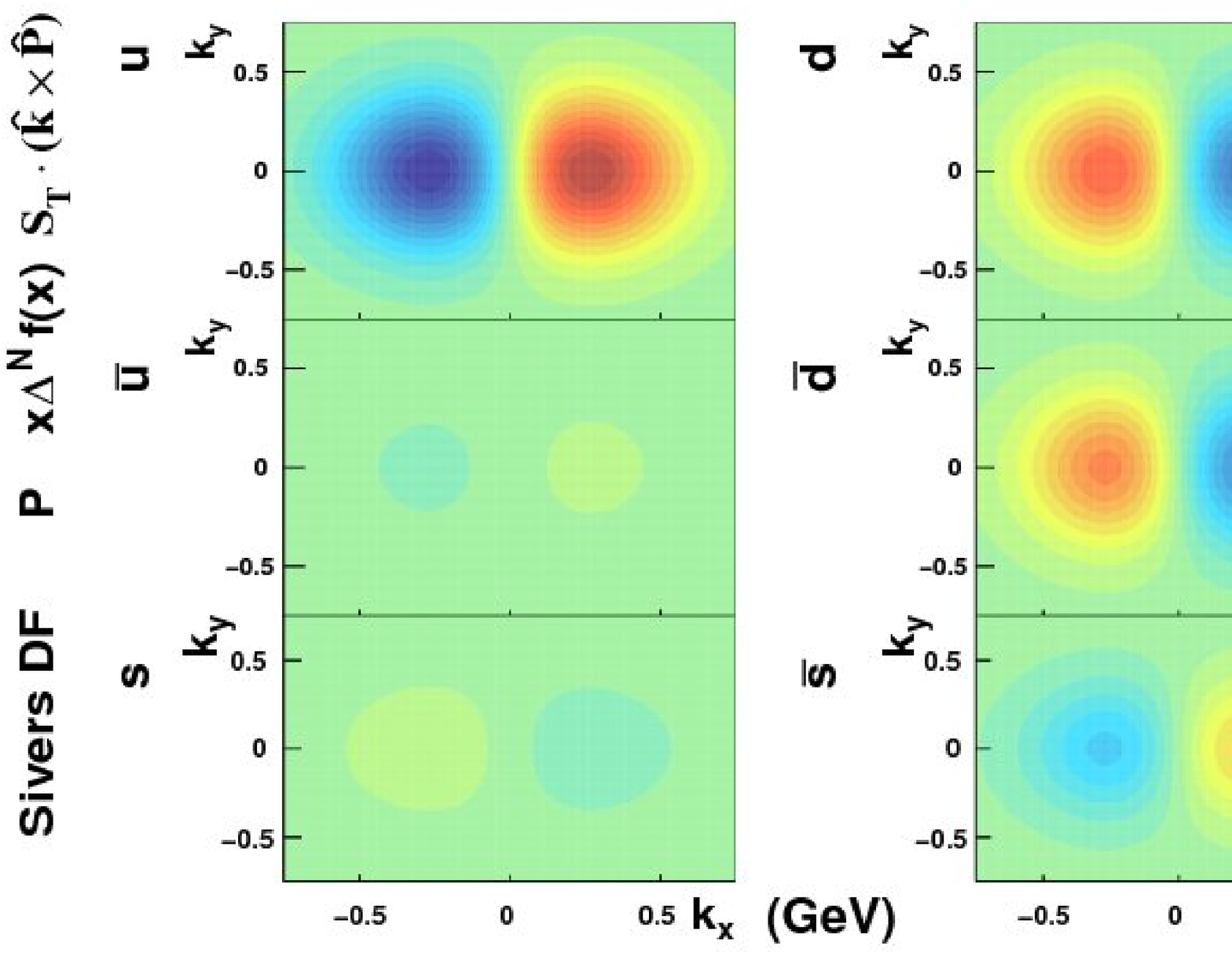}\hskip 3.cm
\includegraphics[width=.35\textwidth,bb= 30 30 740 540]{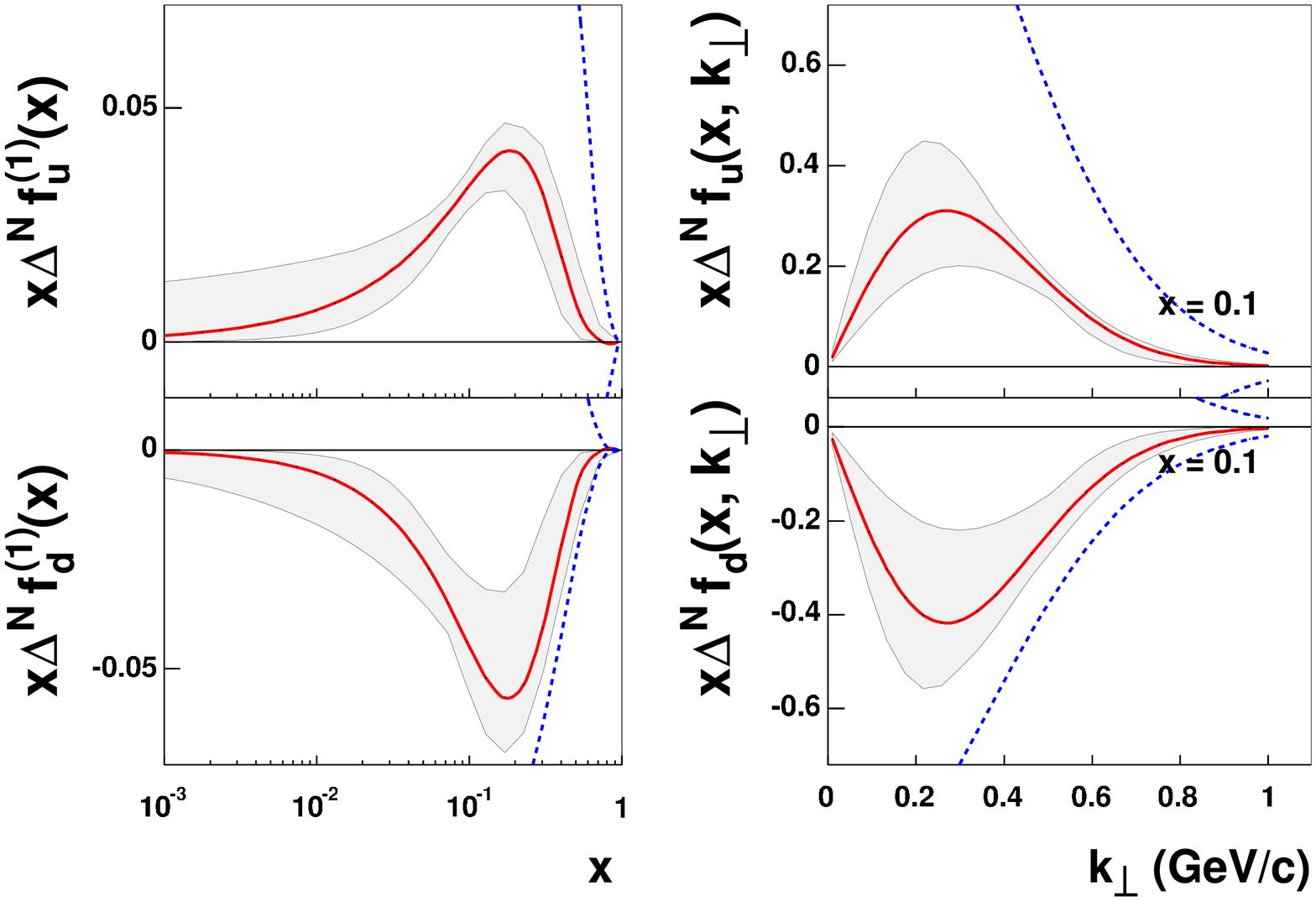}
\caption{Left panel: Sivers function as three dimensional parton distribution.
Right panel: Sivers functions for $u$ and $d$ quarks from extraction of Ref.~\cite{Anselmino:2008sga}.}
\label{fig_2}
\end{figure}

In Fig.~\ref{fig_2} we plot Sivers function extracted from the experimental data from HERMES\cite{Airapetian:2004tw} and COMPASS \cite{Alekseev:2008dn} collaborations, in Fig.~\ref{fig_2} we show a three dimensional parton distribution at $x=0.01$, as can be seen from Fig.~\ref{fig_2} the distribution of partons in a transversely polarised hadron is not rotational symmetric, the distributions have dipole deformation with respect to the ``center'' of the hadron.

\begin{figure}{}
\hskip 0.5cm\includegraphics[width=.25\textwidth,bb= 10 140 540 660,angle=-90]{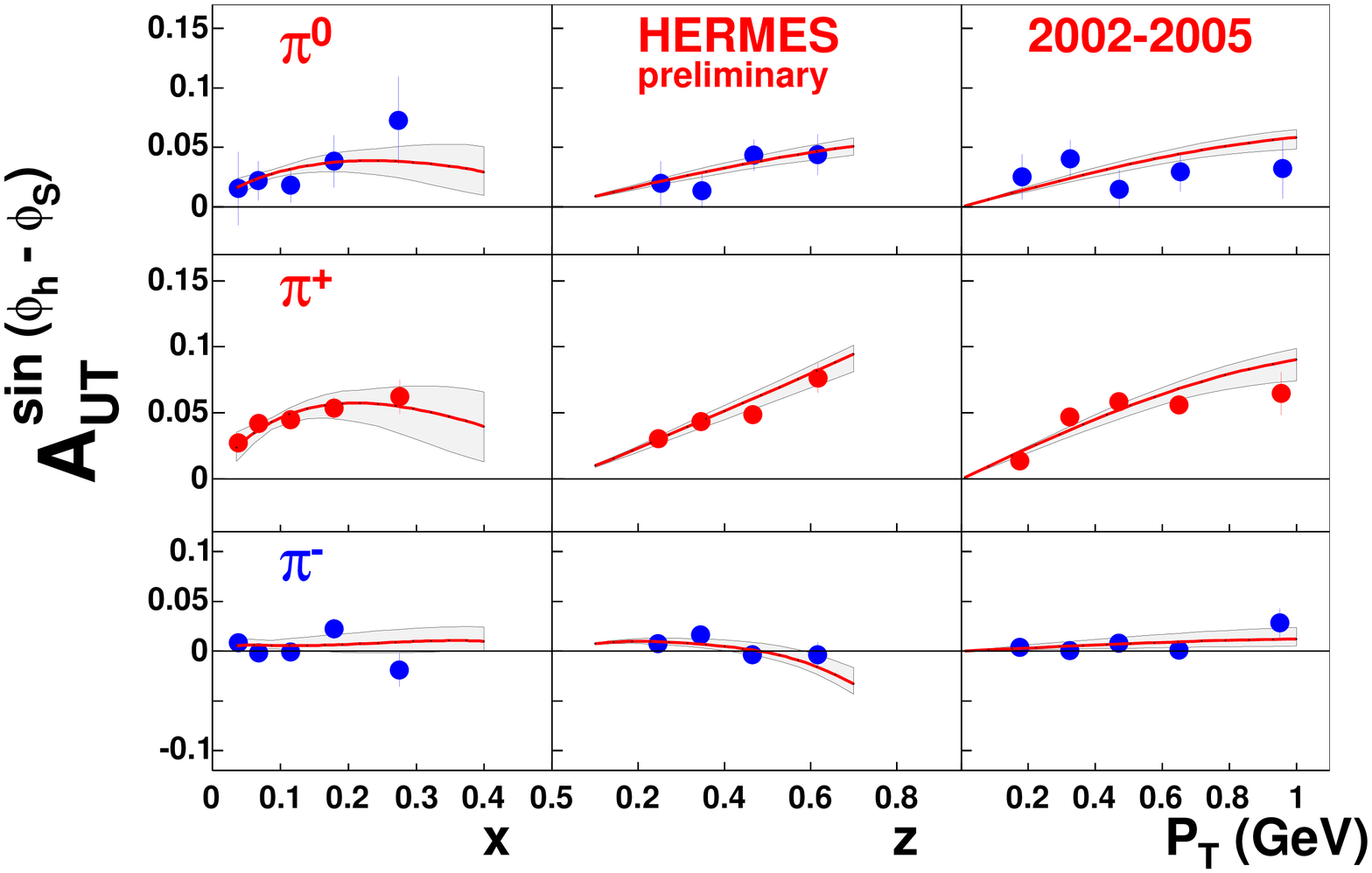}\hskip 2.5cm
\includegraphics[width=.25\textwidth,bb= 10 140 540 660,angle=-90]{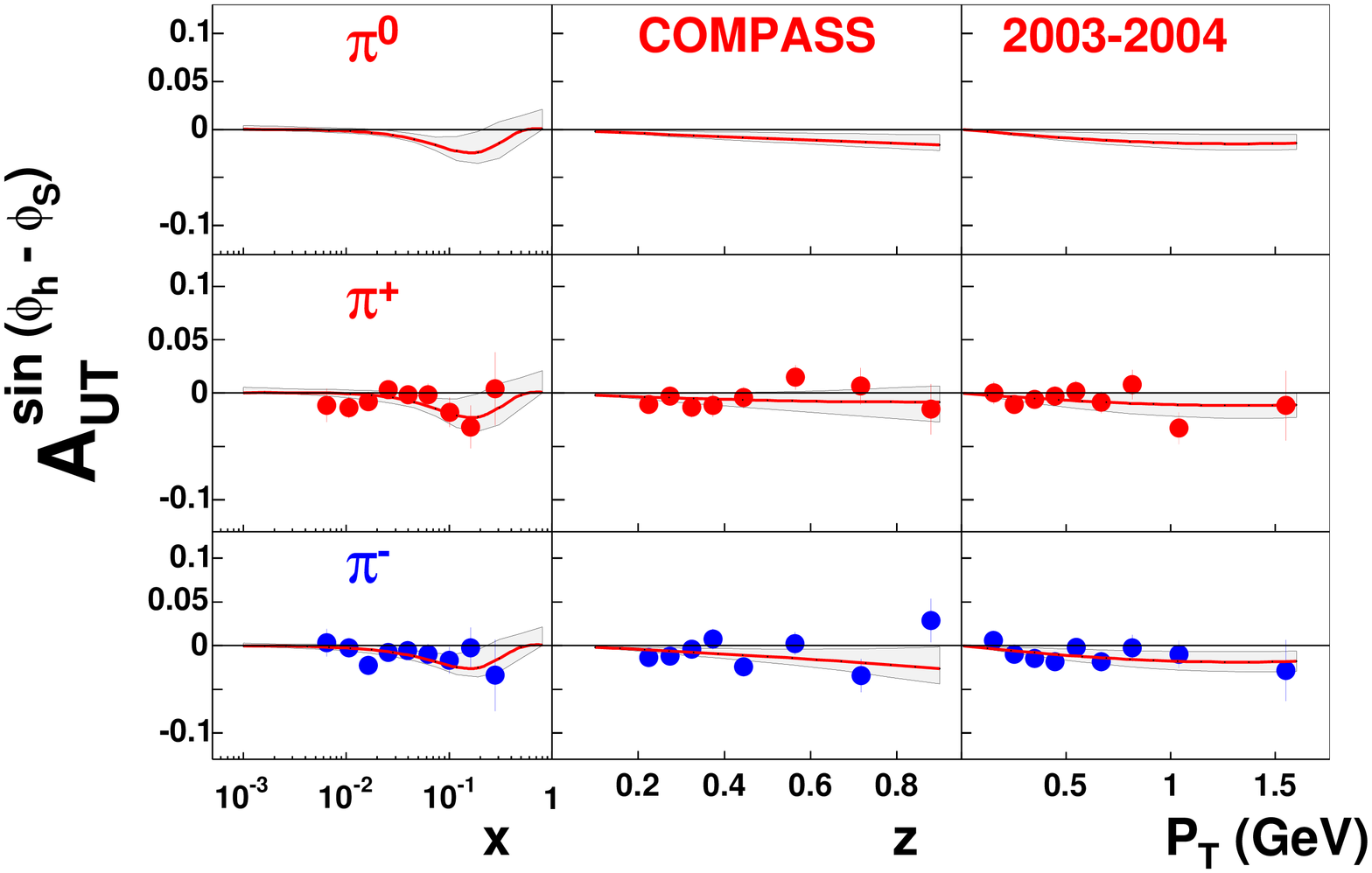}
\caption{Fit\cite{Anselmino:2008sga} of pion HERMES \cite{Airapetian:2004tw} (left panel) and COMPASS \cite{Alekseev:2008dn} data (right panel).}
\label{fig_sivers}
\end{figure}


Fit\cite{Anselmino:2008sga} of HERMES \cite{Airapetian:2004tw} and COMPASS \cite{Alekseev:2008dn} data for $\pi^\pm$ is presented in Fig.~\ref{fig_sivers}.

\noindent {\bf Conclusions}\\
TMDs describe spin structure of the proton. Experimental data from HERMES, COMPASS and BELLE collaborations allow extraction of Sivers function, transversity and Collins fragmentation functions. Future Electron Ion Collider will be a powerful tool to study partonic structure of the nucleon. Three-dimensional picture of partons in the nucleon is one of the main goals of EIC. Flavor decomposition of TMDs will be performed, sea quark and gluon TMDs will be probed and evolution of asymmetries and eventually TMDs themselves will be tested. Large $P_{h\perp}$ range will allow us to study interplay of collinear and TMD factorization schemes. 


\vskip -0.2cm


\end{document}